\documentclass[12pt]{iopart}

\begin{document}

\title{Hydrodynamic equations for a U(N) invariant superfluid}

%

\author{Yi-Cai Zhang$^1$, Shizhong Zhang$^2$}
\address{$^1$ School of Physics and Materials Science, Guangzhou University, Guangzhou 510006, People's Republic of China}
\ead{zhangyicai123456@163.com}
\address{$^2$ Department of Physics and Hong Kong Institute of Quantum Science and Technology, The University of Hong Kong, Hong Kong, China}
\ead{shizhong@hku.hk}

\vspace{10pt}

\begin{abstract}
In this paper, we develop the appropriate set of hydrodynamic equations in a U(N) invariant superfluid that couple the dynamics of superflow and magnetization. In the special case when both the superfluid and normal velocities are zero, the hydrodynamic equations reduce to a generalized version of Landau-Lifshitz equation for ferromagnetism with U(N) symmetry.
When both velocities are non-zero, there appears couplings between the superflow and magnetization dynamics, and the superfluid velocity no longer satisfies the irrotational condition. On the other hand, the magnitude of magnetization is no longer a constant of motion as was the case for the standard Landau-Lifshitz theory. In comparison with the simple superfluid, the first sound and second sounds are modified by a non-zero magnetization through various thermodynamic functions.
For U(2) invariant superfluid,  we get both (zero-) sound wave and a spin wave at zero temperature. It is found that the dispersion of spin wave is always quadratic, which is consistent with detailed microscopic analysis. In the Appendix, we
show that the hydrodynamic equation for a U(N) invariant superfluid can be obtained from the general hydrodynamic equation with arbitrary internal symmetries.
\end{abstract}

%
%
%
%
%

\section{Introduction}
The low-energy and long wave-length dynamics of a many-body system is usually described by hydrodynamics. The relevant hydrodynamic variables typically refer to the densities of conserved quantities, for example, the particle density, the momentum density~\cite{Forster}. In addition, when symmetry breaking is involved, extra thermodynamical variables may arise. For example, in superfluid Helium-4 or a single component Bose-Einstein condensate of atomic Bose gas, the hydrodynamic variable also include the symmetry broken variable, the superfluid velocity $\mathbf{v}_s$ \cite{Chaikin}.

On the other hand, the patterns of symmetry breaking in a multi-component Bose or Fermi system are more complicated and provide a new platform to explore interplay between various different orders. One classic example is that of superfluid Helium three where, depending on the external parameters, different phases are realized with different broken symmetries \cite{Vollhardt}. The hydrodynamical behavior is, as a result, much richer than its Helium-4 counterpart. In cold atom physics, multi-component Bose gases have also attracted a lot of interest because of its various ground states with different magnetic structures. For example, in a spinor-1 Bose-Einstein condensate (BEC), depending on interaction and applied Zeeman field, there may exist ferromagnetic phase, antiferromegnetic phase or polar phase \cite{Ho,Ohmi,Kawaguchi}. Another interesting example is spin-orbit coupled atomic gas~\cite{Zhai,Lin1,zhangshizhong,Yip,zhuqizhong,liyun,Wang2,Cheuk,zhangjinyi,Olson,Khamehchi,Ji}, which hosts plane-wave phase, strip phase, zero-momentum phase. In the last case, the superfluid properties is also very unique with superfluid density greatly suppressed due to enhanced effective mass by spin-obit coupling \cite{normaldensity}, and an anisotropic response in hydrodynamics \cite{Martone2012,Hou2015,Qu2017,zhangyongping,zhangyicai2019}.


In a typical Bose-Einstein condensate of atomic gases, due to the weak inter-atomic interaction, the superfluid behaviors can be described by Gross-Pitaevskii (GP) equation. Furthermore, based on GP equation, by introducing the amplitude and the phase of the condensate wave functions, one can obtain the hydrodynamic description of the system in terms of particle number density $n$ and superfluid velocity $\mathbf{v}_s$ \cite{Pitaevskii, Huibook, Zhengwei2012, zhangyicai2020,Ueda,Andreev2021,ZhangFan2023}. We note, however, that there are several limitations of this method. First of all, the standard GP equation only applies when the interaction is weak and the quantum depletion is negligible. It is much harder to justify the use of GP equation when inter-atomic interaction is strong. Secondly, at finite temperature, the generalization of the GP equation is fairly complicated and the reduction of which to the standard hydrodynamics is not straightforward~\cite{GNZ}. Thirdly, even if we restricted ourselves to the zero-temperature case, a straightforward generalization of the phase-amplitude method to multi-component gas does not bring out the magnetic order explicitly. It is thus useful to construct the low-energy hydrodynamic theory by appealing to symmetry and conservation law directly.


In this paper, we overcome above difficulties by choosing density of conserved quantities as hydrodynamic variables. Using Hamilton method, we develop the hydrodynamic theory for a multi-component superfluid system with $U(N)$ invariant interactions. In this approach, the physical meaning of hydrodynamic variables are clear. In principle, these hydrodynamic equations can be applied at both zero and finite temperatures. We find that due to coupling of superflow and magnetization, the superfluid velocity does not in general satisfy the irrotational condition and the magnitude of magnetization is not a constant in general, as was the case for the standard ferromagnetic theory without damping. Furthermore, based on the hydrodynamic equations obtained,  we calculate the velocities of the first and second sound, and the spin wave excitation in a $U(2)$ invariant superfluid. It is found that both sound velocities are modified by nonzero magnetization.

\section{The model Hamiltonian with a flat band}
To set the stage for our discussion of $U(N)$ superfluid, let us first start with the application of the Hamilton method~\cite{Pokrovskii} in classical liquid to derive the ideal hydrodynamic equations.
From thermodynamics, we know \cite{Chaikin,zhangyicai2019,Shenoy}
\begin{eqnarray}\label{1}
d\epsilon &=Tds+\mu dn+\mathbf{v}\cdot d\mathbf{g},\nonumber\\
f &=\epsilon-Ts-\mathbf{v}\cdot\mathbf{g},\nonumber\\
df &=-sdT+\mu dn-\mathbf{g}\cdot d\mathbf{v},\nonumber\\
p&=-\epsilon +Ts+\mathbf{v}\cdot \mathbf{g}+\mu n,\nonumber \\
d p &= sdT+\mathbf{g}\cdot d\mathbf{v}+nd\mu ,
\end{eqnarray}%
where $\epsilon$ is energy density (energy per unit volume), $f$ is free energy density (free energy per unit volume), $T$ is temperature, $s$ is entropy density (entropy per unit volume), $\mathbf{v}$ is liquid velocity, $\mathbf{g}$ is momentum density (momentum per unit volume), $\mu$ is chemical potential and $n$ is the particle number density (particle number per unit volume). Here the hydrodynamic variables are taken as the number density $n$, momentum density $\mathbf{g}$, and entropy density $s$.
In order to get hydrodynamic equations with Hamilton method, we use the energy density $\epsilon$ to construct an energy functional (Hamiltonian), i.e.,
\begin{eqnarray}\label{2}
H=\int d ^3 \mathbf{r} \epsilon(\mathbf{r}).
\end{eqnarray}
By Eq.(\ref{1}), the Hamilton equation of hydrodynamic variable $A$ is given by
\begin{eqnarray}\label{3}
\frac{\partial A(\mathbf{r}_1)}{\partial t} &=\{A(\mathbf{r}_1),H\}\nonumber\\
&=\int d ^3 \mathbf{r}_2\left[ \frac{\partial\epsilon(\mathbf{r}_2)}{\partial s(\mathbf{r}_2)}\{A,s(\mathbf{r}_2)\}+\frac{\partial\epsilon(\mathbf{r}_2)}{\partial n(\mathbf{r}_2)}\{A,n(\mathbf{r}_2)\}+\frac{\partial\epsilon(\mathbf{r}_2)}{\partial \mathbf{g}(\mathbf{r}_2)}\{A,\mathbf{g}(\mathbf{r}_2)\}\right],\nonumber\\
&=\int d ^3 \mathbf{r}_2\left[T(\mathbf{r}_2)\{A,s(\mathbf{r}_2)\}+\mu(\mathbf{r}_2)\{A,n(\mathbf{r}_2)\}+\mathbf{v}(\mathbf{r}_2)\cdot\{A,\mathbf{g}(\mathbf{r}_2)\}\right],
\end{eqnarray}
where $\{A,B\}$ denotes the classical Poisson's  bracket.

Based on Eqs.(\ref{1}), (\ref{2}) and (\ref{3}), we see that in order to get the equation of motion in terms of hydrodynamic variables, we need to know the commutation relations (Poisson's brackets) among   $n, \mathbf{g}$, and $s$.
In the following, we adopt Landau's method to calculate their commutation relations \cite{Landau,Lifshitz}.
First of all, we know the particle number density $n$ and momentum density can be written in terms of the position and momentum of the particles, i.e.,
\begin{eqnarray}\label{4}
& n(\mathbf{r})=\sum_{i}\delta^{3}(\mathbf{r}_i-\mathbf{r});\nonumber\\
&\mathbf{g}(\mathbf{r})=\sum_{i}\mathbf{p}_i\delta^{3}(\mathbf{r}_i-\mathbf{r}),
\end{eqnarray}
where $\mathbf{r}_i$, $\mathbf{p}_i$ are respectively the position vectors and momentum of $i$-th particle, which form the canonical conjugate pairs, i.e.,
\begin{eqnarray}\label{5}
&\{\mathbf{r}_i,\mathbf{p}_i\}=-\{\mathbf{p}_i,\mathbf{r}_i\}=\delta_{i,j}\nonumber\\
&\{\mathbf{r}_i,\mathbf{r}_j\}=\{\mathbf{p}_i,\mathbf{p}_j\}=0,
\end{eqnarray} and  the classical Poisson's  bracket between $A$ and $B$ is given by
\begin{eqnarray}\label{51}
\{A,B\}=\sum_{k}\left[\frac{\partial A}{\partial\mathbf{ r}_k}\cdot\frac{\partial B}{\partial \mathbf{p}_k}-\frac{\partial A}{\partial \mathbf{p}_k}\cdot\frac{\partial B}{\partial \mathbf{r}_k}\right].
\end{eqnarray}

Using Eqs.(\ref{4}) and (\ref{5}), we find the commutation relations
\begin{eqnarray}
\{n(\mathbf{r}_1),n(\mathbf{r}_2)\} &=0,\label{6-1}\\
\{n(\mathbf{r}_1),\mathbf{g}(\mathbf{r}_2)\} &=n(\mathbf{r}_2)\nabla_2\delta^3(\mathbf{r}_2-\mathbf{r}_1),\label{6-2}\\
\{g_i(\mathbf{r}_1),g_j(\mathbf{r}_2)\} &=g_{i}(\mathbf{r}_2)\nabla_{2j}\delta^3(\mathbf{r}_1-\mathbf{r}_2)-g_j(\mathbf{r}_1)\nabla_{1i}\delta^3(\mathbf{r}_1-\mathbf{r}_2),\label{6-3}
\end{eqnarray}
where $\nabla_2$ stands for ${\partial}/{\partial \mathbf{r}_2}$ and index $i,j=x,y,z$.

In order to get the commutation relations between $s$ and $\mathbf{g}$, $n$, one can make use of their properties under the general coordinate transformation~\cite{Dzyaloshinskii}. It is found that the commutation relations are given by
\begin{eqnarray}
&\{s(\mathbf{r}_1),n(\mathbf{r}_2)\}=\{s(\mathbf{r}_1),s(\mathbf{r}_2)\}=0, \label{8-1}\\
&\{s(\mathbf{r}_1),\mathbf{g}(\mathbf{r}_2)\}=s(\mathbf{r}_2)\nabla \delta^{3}(\mathbf{r}_1-\mathbf{r}_2).\label{8-2}
\end{eqnarray}
Note that the Poisson bracket for entropy density $s(\mathbf{r}_1)$ with momentum density $\mathbf{g}(\mathbf{r}_2)$ are of the same form as that of particle density $n(\mathbf{r}_1)$ with $\mathbf{g}(\mathbf{r}_2)$. This is expected as both $n(\mathbf{r}_1)$ and $s(\mathbf{r}_1)$ are scalar variables that transform in the same way under the general coordinate transformation.


Based on the commutation relations Eqs.(\ref{6-1}-\ref{8-2}), the Hamilton equation (\ref{3}) gives the hydrodynamic equation
\begin{eqnarray}
\frac{\partial n}{\partial t} &=\{n,H\}=-\nabla\cdot (n\mathbf{v}),\label{10-1}\\
\frac{\partial g_j}{\partial t} &=\{g_j,H\}=-n\partial_j \mu-s\partial_j T-g_i\partial_j v_i-\partial_i(v_ig_j),\label{10-2}\\
\frac{\partial s}{\partial t} &=\{s,H\}=-\nabla\cdot (s\mathbf{v}).\label{10-3}
\end{eqnarray}
Moreover, using thermodynamic relations Eq.(\ref{1}),
Eq.(\ref{10-2}) is reduced into the momentum conservation equation, i.e.,
\begin{eqnarray}
\frac{\partial g_j}{\partial t}=\{g_j,H\}=-\partial_i(\pi_{ji}),
\end{eqnarray}%
where the stress tensor
\begin{eqnarray}
\pi_{ji}=p\delta_{ji}+g_j v_i.
\end{eqnarray}%
Furthermore we introduce particle current density $\mathbf{j}$ and velocity $\mathbf{v}$ by
\begin{eqnarray}\label{13}
&\mathbf{j}=\mathbf{g}/m=n\mathbf{v},
\end{eqnarray}
where $m$ is the mass of particle of liquid. In terms of current density $\mathbf{j}$ and velocity $\mathbf{v}$, Eqs.(\ref{10-1}-\ref{10-3}) becomes the standard Euler's equations \cite{fluid}, i.e.,
\begin{eqnarray}
&\frac{\partial n}{\partial t}+\nabla\cdot \mathbf{j}=0,\label{Euler-1}\\
&\frac{\partial \mathbf{v}}{\partial t}+(\mathbf{v}\cdot \nabla )\mathbf{v}=-\frac{ \nabla p}{\rho},\label{Euler-2}\\
&\frac{\partial s}{\partial t}+\nabla\cdot (s\mathbf{v})=0,\label{Euler-3}
\end{eqnarray}
where $\rho=mn$ is mass density (mass per unit volume).
The first equation, Eq.(\ref{Euler-1}), expresses particle number conservation; the second Eq.(\ref{Euler-2}) is the famous Euler's dynamical equation for velocity and the third Eq.(\ref{Euler-3}) is the entropy conservation equation. In the following section, we use a similar method to get the hydrodynamic equation for a superfluid.

%
\section{Hydrodynamics for a simple superfluid }\label{sec:4}
Compare with the classical fluid, superfluid distinguishes itself by the additional thermodynamic variables, the superfluid velocity $\mathbf{v}_s$ that characterize the motion of the superfluid component. As a result, in a superfluid, the whole liquid can be viewed as two different parts, i.e., the normal part with a particle number density $n_n$, the superfluid part with particle number density $n_s$; the total particle density is given by $n=n_n+n_s$. The superfluid component suffers no viscosity and is governed by dynamical laws that is very different from the normal component~\cite{Landau,Lifshitz,Tisza}.
The normal fluid is characterized by a velocity $\mathbf{v}_n$.

For a superfluid system, the free energy density takes the form of \cite{Chaikin,zhangyicai2019}
\begin{eqnarray}
f=f_0(T,n)+\frac{mn\mathbf{v}_{n}^{2}}{2}+\frac{n_s m (\mathbf{v}_s-\mathbf{v}_n)^2}{2}.
\end{eqnarray}%
where $f_0$ is the free energy when liquid is stationary, i.e., $\mathbf{v}_s=\mathbf{v}_n\equiv0$.
The particle number current density $\mathbf{j}$ and the conjugate variable, $\mathbf{h}$, of superfluid velocity $\mathbf{v}_{s}$ are given respectively by  \cite{zhangyicai2019}
\begin{eqnarray}
\mathbf{j} &=-\frac{\partial f}{m\partial \mathbf{v}_{n}}=n_{n}\mathbf{v}_{n}+n_{s}\mathbf{v}_{s} =\mathbf{g}/m \label{h-1} \\
\mathbf{h} &=\frac{\partial f}{\partial \mathbf{v}_{s}}=n_{s}m(\mathbf{v}_{s}-\mathbf{v}_{n})\label{h-2},
\end{eqnarray}%
The  thermodynamic relations of a superfluid takes the form of \cite{zhangyicai2019}
\begin{eqnarray}
d\epsilon &=Tds+\mu dn+\mathbf{v}_{n}\cdot d\mathbf{g}+\mathbf{h}\cdot d\mathbf{v}_{s},\label{relation0-1}\\
f &=\epsilon-Ts-\mathbf{v}_n\cdot\mathbf{g},\label{relation0-2}\\
df &=-sdT+\mu dn-\mathbf{g}\cdot d\mathbf{v}_n+\mathbf{h}\cdot d\mathbf{v}_{s},\label{relation0-3}\\
p &=-\epsilon +Ts+\mathbf{v}_{n}\cdot \mathbf{g}+\mu n, \label{relation0-4}\\
d p &= sdT+\mathbf{g}\cdot d\mathbf{v}_n+nd\mu -\mathbf{h}\cdot d\mathbf{v}_{s}.\label{relation0-5}
\end{eqnarray}%
Similarly, the Hamiltonian is given by
\begin{eqnarray}
H=\int d ^3 \mathbf{r} \epsilon(\mathbf{r}).
\end{eqnarray}
In order to get the equation for $\mathbf{v}_s$,  we need to know the commutation relations between $\mathbf{v}_s$ and other hydrodynamic variables, $n$, $\mathbf{g}$ and $s$.
To this end,  we can either make use of the transformation law of $\mathbf{v}_s$ under the general coordinate transformation~\cite{Dzyaloshinskii} or introduce the Clebsch's representation~\cite{Pokrovskii,Lamb,Lebedev,Lebedev1980} for the momentum density $\mathbf{g}$, i.e,
\begin{eqnarray}\label{7}
\mathbf{g}=n m\mathbf{v}_s+s \nabla \beta
\end{eqnarray}
where
$n$, $s$ are particle number density and entropy density, respectively.
$\mathbf{v}_s$ (superfluid velocity), $\beta$ is the Clebsch variable. Together with $n$, $s$,  they form two independent canonical conjugate pairs with  relations \cite{Pokrovskii}:
\begin{eqnarray}
\{n(\mathbf{r}_1),m\mathbf{v}_s(\mathbf{r}_2)\} &=\nabla \delta^{3}(\mathbf{r}_1-\mathbf{r}_2),\label{19-1}\\
\{s(\mathbf{r}_1),\beta(\mathbf{r}_2)\} &=\delta^{3}(\mathbf{r}_1-\mathbf{r}_2),\label{19-2}\\
\{n(\mathbf{r}_1),\beta(\mathbf{r}_2)\} &=\{s(\mathbf{r}_1),\mathbf{v}_s(\mathbf{r}_2)\}=\{n(\mathbf{r}_1),s(\mathbf{r}_2)\}=0.\label{19-3}
\end{eqnarray}
One can  verify that the above relations Eq.(\ref{7}) and Eqs.(\ref{19-1}-\ref{19-3}) can recover the correct commutation relations Eqs.(\ref{6-1}-\ref{8-2}).

From the Eqs.(\ref{6-1}-\ref{6-3}), Eq.(\ref{7}) and Eqs.(\ref{19-1}-\ref{19-3}), we can get
\begin{eqnarray}
\{\textrm{v}_{si}(\mathbf{r}_1),g_{j}(\mathbf{r}_2)\} &=\textrm{v}_{sj}(\mathbf{r}_2)[\nabla_{2i}\delta^3(\mathbf{r}_2-\mathbf{r}_1)]-\delta^3(\mathbf{r}_2-\mathbf{r}_1)[\nabla_{j}\textrm{v}_{si}-\nabla_{i}\textrm{v}_{sj}],\label{20-1}\\
\{\textrm{v}_{si}(\mathbf{r}_1),\textrm{v}_{sj}(\mathbf{r}_2)\} &=-\frac{1}{mn}\delta^3(\mathbf{r}_2-\mathbf{r}_1)[\nabla_{j}\textrm{v}_{si}-\nabla_{i}\textrm{v}_{sj}].\label{20-2}
\end{eqnarray}

Using commutation relation Eqs.(\ref{6-1}-\ref{8-2}), Eqs.(\ref{19-1}-\ref{20-2}) and thermodynamical relation Eqs.(\ref{relation0-1}-\ref{relation0-5}),
the two-fluid equations can be written in the form of a set of Hamilton equations, i.e.,
\begin{eqnarray}
\frac{\partial n}{\partial t} &=\{n,H\}=-\nabla (n \cdot \mathbf{v}_n)-\nabla \cdot({\mathbf{h}}/{m})=-\nabla \cdot \mathbf{j},\label{26-1} \\
\frac{\partial g_{i}}{\partial t} &=\{g_i,H\}=-\sum_{j}\partial _{j}\pi_{ij},\label{26-2}  \\
\frac{\partial s}{\partial t} &=\{s,H\}=-\nabla\cdot(s\mathbf{v}_n),\label{26-3} \\\nonumber
\frac{\partial \mathbf{v}_{s}}{\partial t} &=\{ \mathbf{v}_{s},H\}\\
&=-\nabla ({\mu}/{m})-\nabla(\mathbf{v}_n \cdot \mathbf{v}_s)
-(\nabla\times \mathbf{v}_s)\times \mathbf{v}_n-\frac{(\nabla\times \mathbf{v}_s)\times \mathbf{h}}{mn} \nonumber\\
&=-\nabla ({\mu}/{m})-\nabla(\mathbf{v}_n \cdot \mathbf{v}_s) -\frac{(\nabla\times \mathbf{v}_s)\times (n_n\mathbf{v}_n+n_s\mathbf{v}_s)}{n}, \label{26-4}
\end{eqnarray}
with the constitutive relations
\begin{eqnarray}
h_i &=mn_s(\textrm{v}_{si}-\textrm{v}_{ni}),\\
j_{i} &=n_{n}\textrm{v}_{ni}+n_{s}\textrm{v}_{si}, \\
g_{i} &=mj_{i}, \\
\pi_{ji} &=p\delta_{ij}+mn_{n}\textrm{v}_{nj}\textrm{v}_{ni}+mn_{s}\textrm{v}_{sj}\textrm{v}_{si}.
\end{eqnarray}
From Eqs.(\ref{26-1}-\ref{26-4}), we see that if initially  the superfluid velocity is irrotational, i.e., $\nabla\times \mathbf{v}_s=0$, then the superfluid velocity would  stay irrotational in the future \cite{Landau}. Therefore, the equation for superfluid velocity becomes:
\begin{eqnarray}
\frac{\partial \mathbf{v}_{s}}{\partial t}+\nabla ({\mu}/{m}+\mathbf{v}_n \cdot \mathbf{v}_s)=0.
\end{eqnarray}
In the following, we shall see that in a muliti-component superfluid with U(N) invariant interactions, due to coupling between superfluid motion and magnetization, the superfluid velocity inevitably has non-vanishing vorticity in general.

\section{Hydrodynamics for a U(N) invariant superfluid}
The general Hamiltonian with $U(N)$-invariant interactions can be written in the following form,
\begin{eqnarray}
H &=\sum_{\sigma}\int dr \psi^{\dagger}_{\sigma}(r)\frac{-\hbar^2{\nabla}^{2}}{2m}\psi_{\sigma}(r) \\\nonumber
&+\frac{1}{2}\sum_{\sigma\sigma'}\int dr_1dr_2\psi^\dagger_{\sigma}(r_1)\psi^\dagger_{\sigma'}(r_2)V^{II}(\mathbf{r}_{1}-\mathbf{r}_{2})\psi_{\sigma'}(r_2)\psi_{\sigma}(r_1)\nonumber\\
&+\frac{1}{2}\sum_{\sigma_1\sigma_2\sigma_3\sigma_4}\int dr_1dr_2T^{a}_{\sigma_1\sigma_2}T^{a}_{\sigma_3\sigma_4} \psi^\dagger_{\sigma_1}(r_1)\psi^\dagger_{\sigma_3}(r_2)V^{SS}(\mathbf{r}_{1}-\mathbf{r}_{2})\psi_{\sigma_4}(r_2)\psi_{\sigma_2}(r_1),\nonumber
\end{eqnarray}%
where $T^a$ is the generator of SU(N) group and $V^{II}$, $V^{SS}$ are density-density and spin-spin interactions, respectively. $\psi^\dagger_{\sigma}(\mathbf{r})$ is a bosonic creation operator that satisfies the the standard communtation relations.

At low temperature, condensation occurs which can be viewed as $\psi_{\sigma}(\mathbf{r})$ acquiring an expectation value in thermal equlibrium. Let us denote it as $\langle\psi\rangle$ which should be viewed as a column vector consisting of elements $\langle\psi_\sigma\rangle$.
Furthermore, it is known that the $U(N)$ group can be viewed as a direct product by U(1) and SU(N), i.e., $U(N)=U(1)\otimes SU(N)$. A general rotation $R$ that acts on order parameter can then be expressed as
\begin{eqnarray}
R\langle\psi\rangle=e^{i\theta^0T^0+i\theta^a T^a}\langle\psi\rangle,
\end{eqnarray}%
where $T^0=I$ is an identity matrix (the generator of U(1) subgroup), and $T^a$ is the generator of SU(N). Phases $\theta^{0}$ and $\theta^{a}$  are the corresponding (real number) rotation angles of general rotations. The corresponding conserved charges are given by \cite{zhangyicai2018}
\begin{eqnarray}
&n(\mathbf{r})=\psi^{\dag}(\mathbf{r})I \psi(\mathbf{r})=\psi^{\dag}(\mathbf{r}) \psi(\mathbf{r}),\\
&n^a(r)=\psi^{\dag}(\mathbf{r}) T^a\psi(\mathbf{r}).
\end{eqnarray}%
The commutation relations among them can be found easily, using $[T^0,T^b]=0$ and $[T^a,T^b]=if^{c}_{ab}T^{c}$ where the structural constant of SU(N) group $f^{c}_{ab}$ are completely antisymmetric with respect to indices ($abc$),
\begin{eqnarray}
&[n(\mathbf{r}_1),n^a(\mathbf{r}_2)] &=0,\label{30-1}\\
&[n^a(\mathbf{r}_1),n^b(\mathbf{r}_2)] &=i f^{c}_{ab} n^{c}(\mathbf{r}_1)\delta^{3}(\mathbf{r}_1-\mathbf{r}_2)\label{30-2},
\end{eqnarray}

In addition, the structure constant $f^{a}_{bc}$ should satisfy  the Jacobi identity \cite{Weinberg}
\begin{eqnarray}\label{Jacobi}
0=f^{\delta}_{\ \alpha\beta}f^{\epsilon}_{\ \delta\gamma}+f^{\delta}_{\ \gamma\alpha}f^{\epsilon}_{\ \delta\beta}+f^{\delta}_{\ \beta\gamma}f^{\epsilon}_{\ \delta\alpha}.
\end{eqnarray}%
For a $U(2)$ invariant superfluid, the system has  four generators, i.e., $T^0=I_{2\times2}$, $T^1=\sigma^x/2$, $T^2=\sigma^y/2$ and $T^3=\sigma^z/2$ (for definiteness, we assume the dimension of representation of U(2) is two, i.e., its fundamental representation). The corresponding conserved charges are particle number $n=\psi^\dagger(\mathbf{r})\psi(\mathbf{r})$, spins (or magnetization) $M^{x,y,z}=\frac{1}{2}\psi^{\dag}(\mathbf{r}) \sigma^{x,y,z}\psi(\mathbf{r})$.
Now their commutation relations are
\begin{eqnarray}
&[n(\mathbf{r}_1),M^a(\mathbf{r}_2)] =0,\nonumber\\
&[M^a(\mathbf{r}_1),M^b(\mathbf{r}_2)] =i\epsilon^{abc}M^{c}(\mathbf{r}_1)\delta^{3}(\mathbf{r}_1-\mathbf{r}_2),
\end{eqnarray}%
where the structural constant $\epsilon^{abc}$ is the Levi-Civita tensor, which is completely antisymmetric with respect to indices ($abc$).

The U(1) rotation generated by $T^0=I$ plays  a special role because it commutes with other generators $T^a$. The spatial derivative of the U(1) phase $\nabla \theta^0$ corresponds to a superfluid motion of the total density with the velocity $\mathbf{v}_s$. For the SU(N) part, we will choose the density of SU(N) conserved charges (generalized spins) $n^a$ as independent variables in constructing our hydrodynamic equation, instead of the phases $\theta^a$ or their derivatives $\nabla\theta^a$. In fact, it can be shown that, under the assumption of a specific form of free energy [see Eq.(\ref{a17}) in the Appendix] that is consistent with the Gross-Pitaevskii analysis, the generalized spin densities suffice to provide a complete description for low-energy hydrodynamics of a U(N) invariant superfluid.

Let us elaborate a bit more on this point. Taking ferromagnetic system as an example, it is known that in the magnetic ordered phase, if a rotation is along magnetization direction, the magnetization would not be affected. However, if the spatial derivatives (gradient) of SU(2) rotation angles are not zero (i.e. there is a gradient of SU(2) Euler angles), such a rotation usually induces a motion of superflow of mass. Furthermore, such a superflow motion can be also described by a superfluid velocity $\mathbf{v}_s$ (see Appendix). Then the superfluid velocity $\mathbf{v}_s$ would in general be a linear combination of the spatial derivatives of U(1) phase and SU(N) rotation angles. As a consequence of this, it is found later that such a superfluid velocity would not satisfy the irrotational condition in general.

On the other hand, if the axis of a rotation is perpendicular to the magnetization direction,
such a motion corresponds to magnetization distortion, which can be described by Landau-Lifshitz equation that only involve magnetization density and its spatial derivatives. Thus we expect that in the U(N) invariant superfluid, the complete hydrodynamic variables can be chosen as the conserved densities $n$, $n^a$, $\mathbf{g}$, $s$, plus a unique superfluid velocity $\mathbf{v}_s$.

In the following, we view the conserved charge density $n$ and $n^a$ as classical mechanical quantities. The quantum commutation relation will be replaced by classical Poisson's brackets, i.e., $[A,B]/i\Rightarrow\{A,B\}$, then the commutation relation in Eqs.(\ref{30-1}-\ref{30-2}) is replaced by
\begin{eqnarray}
&\{n(\mathbf{r}_1),n^a(\mathbf{r}_2)\}=0,\label{28-1}\\
&\{n^a(\mathbf{r}_1),n^b(\mathbf{r}_2)\}=f^{c}_{ab}n^{c}(\mathbf{r}_1)\delta^{3}(\mathbf{\mathbf{r}}_1-\mathbf{r}_2)\label{28-2}
\end{eqnarray}
In comparison with Eqs.(\ref{30-1}-\ref{30-2}), we drop the imagine unit $i$ in the right-hand sides of equations and replace the quantum commutator brackets by classical Poisson's brackets on the left-hand sides.


Similarly, the particle number current  and the conjugate variable of superfluid velocity $\mathbf{v}%
_{s}$ are given respectively by
\begin{eqnarray}
& \mathbf{j}=-\frac{\partial f}{m\partial \mathbf{v}_{n}}=n_{n}\mathbf{%
v}_{n}+n_{s}\mathbf{v}_{s}=\mathbf{g}/m,  \nonumber  \label{h} \\
& \mathbf{h}=\frac{\partial f}{\partial \mathbf{v}_{s}}=n_{s}m(\mathbf{%
v}_{s}-\mathbf{v}_{n}).
\end{eqnarray}%
Since there are now additional conserved charge densities $n^a$, the thermodynamic
relations need to be generalized as
\begin{eqnarray}
&d\epsilon=Tds+\mu dn+\mathbf{v}_{n}\cdot d\mathbf{g}+\mathbf{h}\cdot d%
\mathbf{v}_{s}+\mu^adn^a,\label{566-1}\\
&f=\epsilon-Ts-\mathbf{v}_n\cdot\mathbf{g},\\
&df=-sdT+\mu dn-\mathbf{g}\cdot d\mathbf{v}_n+\mathbf{h}\cdot d%
\mathbf{v}_{s}+\mu^adn^a,\\
& p=-\epsilon +Ts+\mathbf{v}_{n}\cdot \mathbf{g}+\mu n+\mu^{a}n^a,
 \\
&d p= sdT+\mathbf{g}\cdot d\mathbf{v}_n+nd\mu -\mathbf{h}\cdot d%
\mathbf{v}_{s}+n^ad\mu^a,\label{566-5}
\end{eqnarray}%
where  $n^a$ is density for other conserved charges except for particle number.
$\mu^a$ are generalized chemical potential for conserved charge $n^a$.
Similarly, the Hamiltonian for hydrodynamic equation is given by
\begin{eqnarray}
H=\int d ^3 \mathbf{r} \epsilon(\mathbf{r}).
\end{eqnarray}

In order to get the equation of motion of conserved charges $n^a$, we need to know the commutation relations between $n^a$ and  the other independent variables $n$, $\mathbf{g}$, $\mathbf{v}_s$, $s$ which  appear in the above thermodynamic relation Eq.(\ref{566-1}).
First, let us write
\begin{eqnarray}
& n(\mathbf{r})=\sum_{i}\delta^{3}(\mathbf{r}_i-\mathbf{r});\label{281}\\
&n^a(\mathbf{r})=\sum_{i}T^{a}_{i}\delta^{3}(\mathbf{r}_i-\mathbf{r});\label{282}\\
&\mathbf{g}(\mathbf{r})=\sum_{i}\mathbf{p}_i\delta^{3}(\mathbf{r}_i-\mathbf{r}).\label{283}
\end{eqnarray}
From Eqs.(\ref{28-1}-\ref{28-2}) and Eq.(\ref{281}-\ref{283}), we get commutation relations
\begin{eqnarray}\label{36}
& \{n(\mathbf{r}_1),n(\mathbf{r}_2)\}&\equiv 0,\\
& \{n(\mathbf{r}_1),n^a(\mathbf{r}_2)\}&\equiv 0,\\
& \{n(\mathbf{r}_1),\mathbf{g}(\mathbf{r}_2)\}&=n(\mathbf{r}_2)[\nabla_2\delta^{3}(\mathbf{r}_2-\mathbf{r}_1)],\\
&\{n^a(\mathbf{r}_1),n^{b}(\mathbf{r}_2)\}&=f^{c}_{ab}n^{c}(\mathbf{r}_2)\delta^{3}(\mathbf{r}_2-\mathbf{r}_1),\\
& \{n^{a}(\mathbf{r}_1),\mathbf{g}(\mathbf{r}_2)\}&=n^{a}(\mathbf{r}_2)[\nabla_2\delta^{3}(\mathbf{r}_2-\mathbf{r}_1)],\\
&\{g_i(\mathbf{r}_1),g_j(\mathbf{r}_2)\}&=g_{i}(\mathbf{r}_2)\nabla_{2j}\delta^3(\mathbf{r}_1-\mathbf{r}_2)-g_j(\mathbf{r}_1)\nabla_{1i}\delta^3(\mathbf{r}_1-\mathbf{r}_2).
\end{eqnarray}
Similarly, we introduce the superfluid velocity $\mathbf{v}_s$ by
\begin{eqnarray}
\mathbf{g}=n m \mathbf{v}_s+s \nabla \beta,
\end{eqnarray}
and with a straightforward calculation, we get commutation relations for superfluid velocity, i.e.,
\begin{eqnarray}
&[n(\mathbf{r}_1),m\mathbf{v}_{s}(\mathbf{r}_2)]&=\nabla_2\delta^{3}(\mathbf{r}_2-\mathbf{r}_1),\label{38}\\
&[n^a(\mathbf{r}_1),\mathbf{v}_s(\mathbf{r}_2)]&=\frac{n^a(\mathbf{r}_2)}{mn(\mathbf{r}_2)}\nabla_2\delta^{3}(\mathbf{r}_2-\mathbf{r}_1),\\
&[\textrm{v}_{si}(\mathbf{r}_1),g_{j}(\mathbf{r}_2)]&=\{ \textrm{v}_{sj}(\mathbf{r}_2)[\nabla_{2i}\delta^{3}(\mathbf{r}_2-\mathbf{r}_1)]-\delta^{3}(\mathbf{r}_2-\mathbf{r}_1)[\nabla_{j}\textrm{v}_{si}-\nabla_{i}\textrm{v}_{sj}]\};\\
&[\textrm{v}_{si}(\mathbf{r}_1),\textrm{v}_{sj}(\mathbf{r}_2)]&=-\frac{1}{mn}\delta^{3}(\mathbf{r}_2-\mathbf{r}_1)[\nabla_{j}\textrm{v}_{si}-\nabla_{i}\textrm{v}_{sj}],\\
&\{g_i(\mathbf{r}_1),s(\mathbf{r}_2)\}&=s(r_1)\nabla_{2i}\delta^3(\mathbf{r}_1-\mathbf{r}_2),\\
&\{\mathbf{v}_s(\mathbf{r}_1),s(\mathbf{r}_2)\}&=0.\label{38-5}
\end{eqnarray}
In addition, as usual, the commutation between $n^a$ and entropy are zero, i.e.,
\begin{eqnarray}\label{39}
&\{n^a(\mathbf{r}_1),s(\mathbf{r}_2)\}=0.
\end{eqnarray}
This is natural considering that both transform in exactly the same way under general coordinate transformation.

Using the above commutation relations, we can write the equation of motion as follows:
\begin{eqnarray}
&\frac{\partial n(\mathbf{r})}{\partial t}=-\nabla \cdot \mathbf{j},\\
&\frac{\partial g_j}{\partial t}=-\partial_i(\pi_{ji}),\\
&\frac{\partial s}{\partial t}+\nabla \cdot(s\mathbf{v}_n)=0,\\
& \frac{\partial n^{a}(\mathbf{r})}{\partial t}=\{n^{a}(\mathbf{r}),H\}=-\nabla\cdot (n^{a}\mathbf{v}_n)+f^{c}_{ab}n^c\mu^{b}-\nabla \cdot (\frac{n^a\mathbf{h}}{mn}),\label{41}\\
&\frac{\partial \mathbf{v}_{s}}{\partial t}=\{\mathbf{v}_{s},H\}
=-\nabla (\frac{\mu}{m})-\nabla(\mathbf{v}_n \cdot \mathbf{v}_s)-(\nabla\times \mathbf{v}_s)\times \mathbf{v}_n\\
&-\frac{n^a(\nabla \mu^a)}{mn}-\frac{(\nabla\times \mathbf{v}_s)\times \mathbf{h}}{mn}.
\end{eqnarray}

Because $n^a$ is conserved in a U(N) invariant superfluid, its equation can be written in a form of continuity equations.  To guarantee that, we take generalized chemical potential as
\begin{eqnarray}\label{42}
\mu^a=-\alpha \nabla^2 n^a+\mu^{a}_{0},
\end{eqnarray}
where $\alpha$ is a constant, $\mu^{a}_{0}\propto n^a$ is determined by a minimization of free energy when the system is in thermodynamic equilibrium [see Appendix]. We emphasize that this is equivalent to taking the Landau-Lifshitz's energy functional  $\Delta \epsilon\propto \alpha(\nabla M^a)^2$ for magnetization distortions in ferromagnetic system \cite{Lifshitz} .
 By Eq.(\ref{41}) and Eq.(\ref{42}), we get the current $\mathbf{j}^a$ for $n^a$
\begin{eqnarray}
\mathbf{j}^{a}(r)&= \alpha f^{c}_{ab}n^c\nabla n^{b}+n^{a}\mathbf{v}_n+\frac{n^a\mathbf{h}}{mn}\nonumber\\
&=\alpha f^{c}_{ab}n^c\nabla n^{b}+\frac{n^a(n_n\mathbf{v}_n+n_s\mathbf{v}_s)}{n}\nonumber\\
&=\alpha f^{c}_{ab}n^c\nabla n^{b}+\frac{n^a\mathbf{j}}{n}\nonumber\\
&=\alpha f^{c}_{ab}n^c\nabla n^{b}+n^a\mathbf{\bar{v}}.
\end{eqnarray}
In the above equation, we introduce the average velocity $\mathbf{\bar{v}}\equiv \mathbf{j}/n$.
It is shown that the current of SU(N) charge $n^a$ includes two contributions. The first part $\alpha f^{c}_{ab}n^c\nabla n^{b}$ arises from the magnetization distortion, the second part $n^a\mathbf{\mathbf{\bar{v}}}$ comes from non-zero average velocity $\mathbf{\mathbf{\bar{v}}}$.

Finally, we obtain the hydrodynamical equations for the coupling between superfluid motion and magnetization distortion, i.e.,
\begin{eqnarray}\label{continuity}
& \frac{\partial n}{\partial t}+\nabla \cdot \mathbf{j}=0,\\
& \frac{\partial g_{i}}{\partial t}+\sum_{j}\partial _{j}\pi_{ij}=0, \\
& \frac{\partial s}{\partial t}+\nabla\cdot(s\mathbf{v}_n)=0,\\
& \frac{\partial n^a}{\partial t}+\nabla \cdot \mathbf{j}^{a}=0,\label{47}\\
&\frac{\partial \mathbf{v}_{s}}{\partial t}+\nabla (\frac{\mu}{m}+\mathbf{v}_n \cdot \mathbf{v}_s)=-(\nabla\times \mathbf{v}_s)\times \frac{n_n\mathbf{v}_n+n_s \mathbf{v}_s}{n}-\frac{n^a(\nabla \mu^a)}{mn}, \label{continuity11}
\end{eqnarray}
with again the constitutive relations
\begin{eqnarray}
j_{i} &=n_{n}\textrm{v}_{ni}+n_{s}\textrm{v}_{si}, \\
g_{i} &=mj_{i}, \\
\pi _{ji} &=p\delta _{ij}+mn_{n}\textrm{v}_{nj}\textrm{v}_{ni}+mn_{s}\textrm{v}_{sj}\textrm{v}_{si}\\
\mathbf{j}^{a} &=\alpha f^{c}_{ab}n^c\nabla n^{b}+n^{a}\mathbf{v}_n+\frac{n^a\mathbf{h}}{mn}\nonumber\\
&=\alpha f^{c}_{ab}n^c\nabla n^{b}+\frac{n^a(n_n\mathbf{v}_n+n_s\mathbf{v}_s)}{n},\nonumber\\
&=\alpha f^{c}_{ab}n^c\nabla n^{b}+n^a\mathbf{\bar{v}}.
\end{eqnarray}
The above equations (\ref{continuity}-\ref{continuity11}) are the main results of this paper. In Appendix, we show that the same set of equations can  be also derived from a general hydrodynamical equation with an arbitrary internal symmetry group. The above equations have several important characteristics that is worth emphasizing.

First, when $\mathbf{v}_n=0$ and  $\textbf{v}_s=0$, the equation for $n^a$ reduces to generalized Landau-Lifshitz's equation in ferromagnetization theory. For general cases, the superfluid motions and magnetization motions are coupled together.

Second, we see that the superfluid velocity is not irrotational due to source term $-\frac{n_a(\nabla \mu_a)}{mn}$, which arising from the coupling of magnetization. Furthermore, it is found that, different from the usual two-fluid case, here the superfluid velocity is a mixture of several velocities defined by both the U(1) and SU(N) groups [see Eq.(\ref{a242}) in the Appendix].
The breaking of irrotationality arises from the mixture of several velocities. A similar phenomenon where superfluid velocity does not satisfy the irrotational condition also appears in spinor-1 Bose-Einstein condensate \cite{Ueda} and in spin-orbit coupled Bose gases~\cite{Stringari2017}.


Third,  the motion of magnetization is also affected by superfluid motion. Due to the coupling between magnetization and superfluid motions, the magnitude of magnetization $M=\sqrt{n^an^a}$ is no longer a constant of motion, i.e, $n^a\partial_tn^a\neq0$.
 This is because by Eq.(\ref{47}), we get
 \begin{eqnarray}
\frac{\partial (n^an^a)}{\partial t}=2n^a\partial_tn^a=-2n^a\nabla\cdot\left[\frac{n^a\mathbf{j}}{n}\right]=-2n^a\nabla\cdot (n^a\mathbf{\bar{v}})\neq0
\end{eqnarray}
which is different from the usual Landau-Lifshitz theory where the the magnitude of magnetization is always a constant.

\section{Sound waves and spin waves}
Having established the general hydrodynamic equation for a U(N) invariant superfluid that applies to both boson and fermion systems, irrespective of the strength of inter-atomic interaction, let us investigate its possible collective excitations, including sound wave and spin wave. When the oscillation amplitudes are small, we can neglect the second order terms in velocities of Eq.(\ref{continuity})-(\ref{continuity11}), i.e,
\begin{eqnarray}
& \frac{\partial n}{\partial t}+\nabla \cdot \mathbf{j}=0,
\label{continuityaa} \\
& \frac{\partial g_{i}}{\partial t}+\nabla _{i}p=0. \\
& \frac{\partial s}{\partial t}+\nabla\cdot(s\mathbf{v}_n)=0.\\
& \frac{\partial n^a}{\partial t}+\nabla \cdot \mathbf{j}^{a}=0.\\
&\frac{\partial \mathbf{v}_{s}}{\partial t}+\nabla (\frac{\mu}{m})=-\frac{n_a(\nabla \mu_a)}{mn}
\end{eqnarray}%
with constitutive relations
\begin{eqnarray*}
& j_{i}=n_{n}\textrm{v}_{ni}+n_{s}\textrm{v}_{si}=g_i/m, \\
&\mathbf{j}^{a}=\alpha f^{c}_{ab}n^c\nabla n^{b}+\frac{n^a(n_n\mathbf{v}_n+n_s\mathbf{v}_s)}{n}
\end{eqnarray*}

After introducing the entropy per unit mass, i.e., $\tilde{s}=s/(mn)$ and using thermodynamic relation Eqs.(\ref{566-1}-\ref{566-5}), we get
\begin{eqnarray}\label{56}
& \frac{\partial^2 n}{\partial t^2}=\frac{\nabla^2 p}{m},\\
& \frac{\partial^2 \tilde{s}}{\partial t^2}=\frac{n_s\tilde{s}^2}{n_n}\nabla^2T, \\
& \frac{\partial n^a}{\partial t}+\alpha f^{c}_{ab}n^c\nabla^2n^b-\frac{n^a}{n}\frac{\partial n}{\partial t}=0.
\end{eqnarray}%
Using the thermodynamic relation Eqs.(\ref{566-1}-\ref{566-5}), once we know the free energy, we can get the first, second sounds and spin wave.

We take U(2) invariant superfluid as an example and furthermore we assume that when the system is in thermodynamic equilibrium,  $n^{x(y)}=\bar{m}_x=\bar{m}_y=0$, $n=\bar{n}\neq0$, $n^z=\bar{m}_z\neq0$. The above equations reduce into
\begin{eqnarray}\label{57}
& \frac{\partial^2 n}{\partial t^2}=\frac{\nabla^2 p}{m},\label{57}\\
& \frac{\partial^2 \tilde{s}}{\partial t^2}=\frac{n_s\tilde{s}^2}{n_n}\nabla^2T, \\
& \frac{\partial m_x}{\partial t}+\alpha \bar{m}_z\nabla^2m_y=0,\\
& \frac{\partial m_y}{\partial t}-\alpha \bar{m}_z\nabla^2m_x=0,\\
& \frac{\partial m_z}{\partial t}-\frac{\bar{m}_z\partial n}{\bar{n}\partial t}=0.\label{57-4}
\end{eqnarray}
Choosing $n$, $n^a$ and $\tilde{s}$ as independent variables, we have
\begin{eqnarray}
dp=&\left.\frac{\partial p}{\partial n}\right|_{m_{xyz},\tilde{s}} dn+\left.\frac{\partial p}{\partial \tilde{s}}\right|_{n,m_{xyz}} d\tilde{s}+\left.\frac{\partial p}{\partial m_x}\right|_{n,\tilde{s},m_{yz}} d\tilde{m_x}
\nonumber\\
&+\left.\frac{\partial p}{\partial m_y}\right|_{n,\tilde{s},m_{xz}} d\tilde{m_y}+\left.\frac{\partial p}{\partial m_z}\right|_{n,\tilde{s},m_{yx}} d\tilde{m_z},\label{58-1}\\
dT=&\left.\frac{\partial T}{\partial n}\right|_{m_{xyz},\tilde{s}} dn+\left.\frac{\partial T}{\partial \tilde{s}}\right|_{n,m_{xyz}} d\tilde{s}+\left.\frac{\partial T}{\partial m_x}\right|_{n,\tilde{s},m_{yz}} d\tilde{m_x}
\nonumber\\
&+\left.\frac{\partial T}{\partial m_y}\right|_{n,\tilde{s},m_{xz}} d\tilde{m_y}+\left.\frac{\partial T}{\partial m_z}\right|_{n,\tilde{s},m_{yx}} d\tilde{m_z},\label{58-2}
\end{eqnarray}
Substituting Eqs.(\ref{58-1}-\ref{58-2}) into Eq.(\ref{57}-\ref{57-4}), we obtain two sound (the first and second sound) waves $\omega_{\pm}({\bf q})=c_{\pm} |{\bf q}|$ and one spin wave, $\omega_{\rm spin}({\bf q})=\alpha \bar{m}_z |{\bf q}|^2$. Here the sound velocities are given by
\begin{equation}
c_{\pm}=\frac{1}{\sqrt{2\bar{n}}}(A\pm \sqrt{\Delta})^{1/2}\label{cpm}
\end{equation}
with
\begin{eqnarray}
    A &= (a_{11}+a_{22})\bar{n}+a_{15}\bar{m}_z\nonumber\\
    B &= a_{15}^{2}\bar{m}_{z}^2+2a_{15}(a_{11}-a_{22})\bar{n}\bar{m}_z\nonumber\\
    &+\bar{n}[4a_{12}a_{25}\bar{m}_z+(4a_{12}a_{21}+(a_{11}-a_{22})^2)\bar{n}]\nonumber
\end{eqnarray}
where the coefficients $a_{ij}$'s are given by
\begin{eqnarray}\label{60}
& a_{11}=\left.\frac{\partial p}{m\partial n}\right|_{m_{xyz},\tilde{s}}; \ a_{12}=\left.\frac{\partial p}{m\partial \tilde{s}}\right|_{m_{xyz},n}; \ a_{15}=\left.\frac{\partial p}{m\partial m_z}\right|_{m_{xy},n,\tilde{s}};\nonumber\\
& a_{21}=\left.\frac{n_s\tilde{s}^{2}}{n_n}\frac{\partial T}{\partial n}\right|_{m_{xyz},\tilde{s}}; \ a_{22}=\left.\frac{n_s\tilde{s}^{2}}{n_n}\frac{\partial T}{\partial \tilde{s}}\right|_{m_{xyz},n}; \ a_{25}=\left.\frac{n_s\tilde{s}^{2}}{n_n}\frac{\partial T}{\partial m_z}\right|_{m_{xy},n,\tilde{s}}.
\end{eqnarray}
with all other $a_{ij}=0$. In the above equations, all the physical quantities take their values at thermodynamic equilibrium. We see that in comparison with a simple superfluid, the first and second sound have been changed by the magnetization in U(N) invariant superfluid trough thermodynamic function $p$ and $T$. In the case of a simple superfluid, $m_z=a_{15}=a_{25}\equiv0$, the first and second sounds are given by
\begin{eqnarray}
& \omega_1=c_1 q=\frac{\sqrt{(a_{11}+a_{22})+\sqrt{4a_{12}a_{21}+(a_{11}-a_{22})^2}}}{\sqrt{2}}q,\\
& \omega_2=c_2 q=\frac{\sqrt{(a_{11}+a_{22})-\sqrt{4a_{12}a_{21}+(a_{11}-a_{22})^2}}}{\sqrt{2}}q.
\end{eqnarray}

At zero temperature $T=0$, $\tilde{s}=0$, $a_{21}=a_{22}=a_{25}=0$, by Eq.(\ref{cpm}), we get only a zero sound and a spin wave, i.e.,
\begin{eqnarray}\label{59}
& \omega_0(|{\bf q}|)=\omega_\pm(|{\bf q}|)=\sqrt{a_{11}+\frac{a_{15}\bar{m}_{z}}{\bar{n}}}|{\bf q}|=\sqrt{\left.\frac{\partial p}{m\partial n}\right|_{m_{xyz}}+\left.\frac{\bar{m}_z\partial p}{m \bar{n}\partial m_z}\right|_{m_{xy},n}}q,\nonumber\\
& \omega_{\rm spin}=\alpha \bar{m}_zq^2,
\end{eqnarray}
Near the ground state, the spin-wave dispersion is quadratic, which is consistent with detailed microscopic analysis \cite{zhangyicai2018}.

Furthermore, for weakly interacting atomic gas,  the ground state energy density is given by
\begin{equation}
    \epsilon=\frac{1}{2}g_0\bar{n}^2+\frac{1}{2}g_2\bar{n}_{a}^{2}=\frac{1}{2}g_0n^2+\frac{1}{2}g_2\bar{m}_{z}^{2}
\end{equation}
where $g_0$ is density-density interaction parameter and $g_2$ is the spin-spin interaction parameter. Then by Eqs.(\ref{566-1}-\ref{566-5}), $p=\frac{1}{2}g_0n^2+\frac{1}{2}g_2\bar{m}_{z}^{2}$ and the sound velocity is reduced to
\begin{eqnarray}
c_{0}=\sqrt{\left.\frac{\partial p}{m\partial n}\right|_{m_{xyz}}+\left.\frac{\bar{m}_z\partial p}{m \bar{n}\partial m_z}\right|_{m_{xy},n}}=\sqrt{\frac{g_0 n}{m}+\frac{g_2 \bar{m}_{z}^{2}}{m\bar{n}}}
\end{eqnarray}%
We get a linear dispersion phonon $\omega_0(q)=c_0q$ with sound velocity $c_1=\sqrt{\frac{g_0 n}{m}+\frac{g_2 \bar{m}_{z}^{2}}{m\bar{n}}}$ and a quadratic dispersion spin wave $\omega_{\rm spin}(q)=\alpha \bar{m}_{z}q^2$. We see the spin-spin interaction also has contribute in the sound velocity when a nonzero magnetization appears.

\section{Conclusions}\label{sec:5}
In summary, we generalize the hydrodynamic equation to a superfluid system with U(N) invariant interactions. Generally speaking, the superfluid motion and magnetization motion are coupled together. When the superfluid is stationary, our equations can be reduced to Landau-Lifshitz equation in ferromagnetization theory. Due to the coupling of superfluid motion and magnetization distortion, the superfluid velocity does not satisfy the irrotational condition and the magnitude of magnetization is no longer a constant. In addition, it is found that a non-zero magnetization modifies the sound velocities through thermodynamic relations.

\section{Acknowledgments}
This work was supported by the NSFC under Grants Nos. 11874127, the Joint Fund with
Guangzhou Municipality under No. 202201020137, and the Starting Research Fund from
Guangzhou University under Grant No.
RQ 2020083. SZ is supported by HK GRF Grants No. 17304820 and No. 17304719, CRF Grants No. C6009-20G and No. C7012-21G, and a RGC Fellowship Award No. HKU RFS2223-7S03.

\section{Appendix}

In this Appendix, we review the main results in references \cite{Lebedev,Lebedev1980} which provide a general formalism for hydrodynamics in a superfluid  with arbitrary internal symmetry.
Based on the general hydrodynamical equations  and an assumption of free energy of a U(N) invariant superfluid, we can re-derive the hydrodynamical equations Eq.(\ref{continuity})-Eq.(\ref{continuity11}) for a superfluid with U(N) invariant interactions in the main text.

\subsection{The hydrodynamics with arbitrary symmetry group }
The basic idea in \cite{Lebedev,Lebedev1980} is to define a set of physical quantities for hydrodynamical description and then calculate their commutation relations. Equations of motions are then obtained from Hamilton equations. Let us thus first consider the variations of order parameter for multi-component system (to simplify the notation, we use $\psi$ to denote a column vector of $\psi_a$):
  \begin{eqnarray}\label{a1}
  \delta \psi=iT^0\delta\theta^0\psi+iT^a\delta\theta^a\psi=i\delta\theta^0\psi+iT^a\delta\theta^a\psi.
\end{eqnarray}%
In the above equation, we have neglected the variations of amplitude of order parameter $\psi$. The superfluid velocities $\mathbf{v}_{s}^0$ and $\mathbf{v}_{s}^a$, and the scalar function $\Omega^{a}$ are defined by
 \begin{eqnarray}
  &\nabla \psi=i\mathbf{v}_{s}^0\psi+iT^a\mathbf{v}_{s}^a\psi,\label{a2-1}\\
  &\partial_t \psi=i\Omega^0\psi+iT^a\Omega^a\psi.\label{a2-2}
\end{eqnarray}%
By Eqs.(\ref{a2-1}-\ref{a2-2}) and $\partial_t(\nabla \psi)=\nabla(\partial_t \psi)$, we obtain
the equation of motion for $\mathbf{v}^a$
\begin{eqnarray}
&\frac{\partial \textrm{v}_{si}^0}{\partial t}=-\nabla_i\Omega^0,\label{a41}\\
&\frac{\partial \textrm{v}_{si}^a}{\partial t}=-\nabla_i\Omega^a+f^a_{\ bc}\Omega^b \textrm{v}^{c}_{si}.\label{a42}
\end{eqnarray}%
 In the following, we will determine  the specific function form of $\Omega^0$ and $\Omega^a$ [see Eq.(\ref{a15}) below].

From Eqs.(\ref{a2-1}-\ref{a2-2}), we get
 \begin{eqnarray}
0 &= \nabla\times\nabla \psi=i\nabla\times \mathbf{v}_{s}^aT^a\psi+(-i)(i)T^a\mathbf{v}_{s}^a\times\nabla\psi\nonumber\\
 &=i\nabla\times \mathbf{v}_{s}^aT^a\psi+\frac{1}{2}\mathbf{v}_{s}^a\times \mathbf{v}_{s}^{b}[T^aT^b-T^bT^a]\psi\nonumber\\
 &=i\nabla\times \mathbf{v}_{s}^cT^c\psi+i\frac{1}{2}\mathbf{v}_{s}^a\times \mathbf{v}_{s}^{b}f^{c}_{ \ ab}T^c\psi=0,
\end{eqnarray}%
that is, because $\psi$ is arbitrary
\begin{eqnarray}\label{zerocurvature}
&\nabla\times \mathbf{v}_{s}^c=-\frac{1}{2}f^{c}_{ \ ab}\mathbf{v}_{s}^a\times \mathbf{v}_{s}^{b}.
\end{eqnarray}
Or in component form
\begin{eqnarray}
\quad \nabla_i \textrm{v}_{sj}^c-\nabla_j \textrm{v}_{si}^c+f^{c}_{ \ ab}\textrm{v}_{si
}^a \textrm{v}_{sj}^{b}=0.
\end{eqnarray}%
If we interpret $\mathbf{v}_{s}^{a}$ as gauge potential $\mathbf{A}^{a}$, then Eq.(\ref{zerocurvature}) is the zero-curvature condition for $\mathbf{v}_{s}^{a}$, i.e.,
 \begin{eqnarray}
 F^{c}_{i,j}T^c\phi=[D_i,D_j]\phi=[\nabla_i A_{sj}^c-\nabla_j A_{si}^c+f^{c}_{ \ ab}A_{si
}^a A_{sj}^{b}]T^c\phi=0
\end{eqnarray} where $D_i=\nabla_i-iA^{a}_i T^a$
is the covariant derivative, $F$ is curvature, $\phi$ is matter field in the standard Yang-Mills theory.


In order to get the commutation relations, we promote the order parameter $\psi$ as quantum mechanical field operator. Its hermitian conjugate is $\psi^{\dag}$, satisfying quantum commutation brackets, i.e.,
\begin{eqnarray}
&[\psi_\mu(\mathbf{r}_1), \psi^{\dag}_\nu(\mathbf{r}_2)]=\delta^3(\mathbf{r}_1-\mathbf{r}_2)\delta_{\mu,\nu},\label{a81-1}\\
&[\psi_\mu(\mathbf{r}_1), \psi_\nu(\mathbf{r}_2)]=[\psi^{\dag}_\mu(\mathbf{r}_1), \psi^{\dag}_\nu(\mathbf{r}_2)]=0.\label{a81-2}
\end{eqnarray}%
The particle number density $n$ and conserved  charge  (generalized spin)  density $n^a$ \cite{zhangyicai2018} are given by
\begin{eqnarray}\label{a91}
n=\psi^{\dag}_\mu\psi_\mu; \ n^a=\psi^{\dag}_\mu T^a_{\mu,\nu}\psi_{\nu}.
\end{eqnarray}%
The momentum density is (setting $\hbar=m=1$)
\begin{eqnarray}\label{a9}
\mathbf{g}(\mathbf{r})=-i\frac{\hbar}{m}\psi^{\dag}_\mu(\mathbf{r})\nabla\psi_\mu(\mathbf{r})=-i\psi^{\dag}_\mu(\mathbf{r})\nabla\psi_\mu(\mathbf{r}).
\end{eqnarray}%
In the above equation, the repeated indices are summed.

Now the thermodynamic relations can be written as
\begin{eqnarray}
d\epsilon &=Tds+\textrm{v}_{nj}dg_j+\mu^0dn+h^0_{j}d\textrm{v}^{0}_{sj}+\mu^adn^a+h^a_{j}d\textrm{v}^{a}_{sj}\label{a101-1}\\
f &=\epsilon-Ts-\mathbf{v}_n\cdot\mathbf{g},\\
df &=-sdT+\mu dn-\mathbf{g}\cdot d\mathbf{v}_n+\mathbf{h}^0\cdot d%
\mathbf{v}^{0}_{s}+\mathbf{h}^a\cdot d%
\mathbf{v}^{a}_{s}+\mu^adn^a,\\
p &=-\epsilon+Ts+\textrm{v}_{nj}g_j+\mu^0n+\mu^an^a\\
dp &=sdT+\mathbf{g}\cdot d\mathbf{v}_n+nd\mu^0+n^ad\mu^a-h^0_{j}d\textrm{v}^{0}_{sj}-h^a_{j}d\textrm{v}^{a}_{sj}.\label{a101-5}
\end{eqnarray}%
From Eq.(\ref{a2-1}), we can see that definitions of superfluid velocities only involve $\psi$, rather than $\psi^{\dag}$. So we can assume they commutate each other, i.e.,
\begin{eqnarray}
  [\mathbf{v}^a_{s}(\mathbf{r}_1),\mathbf{v}^{b}_{s}(\mathbf{r}_2)]=0.
\end{eqnarray}%

Using Eq.(\ref{a2-1}) and Eqs.(\ref{a81-1}-\ref{a9}),  the commutation relations can be obtained straightforward,  i.e.,
\begin{eqnarray}\label{a123}
  &[n^{a}(\mathbf{r}_1),n^{b}(\mathbf{r}_2)]=if^{c}_{\ ab}n^{c}(\mathbf{r}_1)\delta(\mathbf{r}_1-\mathbf{r}_2),\label{a123}\\
  &[n^{a}(\mathbf{r}_1),\psi(\mathbf{r}_2)]=-\delta^{3}(\mathbf{r}_1-\mathbf{r}_2)T^{a}\psi(\mathbf{r}_1),\label{a123-1}\\
  &[n^{a}(\mathbf{r}_1),\textrm{v}^{b}_{si}(\mathbf{r}_2)]=i\delta_{ab}\nabla_{2i}\delta^3(\mathbf{r}_1-\mathbf{r}_2)-if^{b}_{\ ac}\textrm{v}^{c}_{s}(\mathbf{r}_1)\delta^{3}(\mathbf{r}_1-\mathbf{r}_2),\\
  &[n^{a}(\mathbf{r}_1),g_{i}(\mathbf{r}_2)]=in^{a}(\mathbf{r}_2)\nabla_{2i}\delta^{3}(\mathbf{r}_1-\mathbf{r}_2),\\
  &[g_{1i},g_{2j}]=ig_{2j}\nabla_{2i}\delta^{3}(\mathbf{r}_1-\mathbf{r}_2)+i[\nabla_ig_j]\delta^{3}(\mathbf{r}_1-\mathbf{r}_2)+ig_{2i}\nabla_{2j}\delta^{3}(\mathbf{r}_1-\mathbf{r}_2),\\
  &[g_{1i},\textrm{v}^{a}_{2j}]=i\textrm{v}^{a}_{2si}\nabla_{2j}\delta^{3}(\mathbf{r}_1-\mathbf{r}_2)+i[\nabla_j\textrm{v}^{a}_{si}]\delta^{3}(\mathbf{r}_1-\mathbf{r}_2)\nonumber\\
  &+i[\nabla_i\textrm{v}^{a}_{sj}-\nabla_j\textrm{v}^{a}_{si}]\delta^{3}(\mathbf{r}_1-\mathbf{r}_2),\label{a123-5}
\end{eqnarray}%
where $g_{1i}$ stands for $g_{i}(\mathbf{r}_1)$.
Similarly to that in the main text, we can write
\begin{eqnarray}%
&[g_i(\mathbf{r}_1),s(\mathbf{r}_2)]=is(\mathbf{r}_1)\nabla_2\delta(\mathbf{r}_1-\mathbf{r}_2),\\
&[s(\mathbf{r}_1),n^a(\mathbf{r}_2)]=[s(\mathbf{r}_1),\mathbf{v}_{s}^a(\mathbf{r}_2)]=0.
\end{eqnarray}%

Now, with the replacement  $[A,B]/i\Rightarrow\{A,B\}$, the equations of motions are
\begin{eqnarray}
\frac{\partial n}{\partial t} &=-\nabla_i[n\textrm{v}_{ni}+h^{0}_{i}],\label{a151}\\
\frac{\partial n^a}{\partial t} &=-\nabla_i[n^a\textrm{v}_{ni}+h^{a}_{i}]+f^{a}_{\ bc}[\mu^b n^c+h^{b}_{i}\textrm{v}^{c}_{si}]\label{a151-1},\\
\frac{\partial \textrm{v}^0_{si}}{\partial t} &=-\nabla_i[\mu^0+\mathbf{v}^{0}_{s}\cdot \mathbf{v}_{n}],\label{a151-2}\\
\frac{\partial \textrm{v}^a_{si}}{\partial t} &=-\nabla_i[\mu^a+\mathbf{v}^{a}_{s}\cdot \mathbf{v}_{n}]+f^{a}_{\ bc}[\mu^b+\mathbf{v}^{b}_{s}\cdot \mathbf{v}_{n}]\textrm{v}^{c}_{si},\label{a151-3}\\
\frac{\partial g_{i}}{\partial t} &=-\nabla_j\pi_{ij},\label{a151-4}\\
\frac{\partial s}{\partial t}&=-\nabla_i (s\textrm{v}_{ni})\label{a151-5},
\end{eqnarray}%
where $\pi_{ij}=p\delta_{ij}+g_i\textrm{v}_{nj}+\textrm{v}^{0}_{si}h^{0}_{j}+\textrm{v}^{a}_{si}h^{a}_{j}$.
Comparing it with Eqs.(\ref{a41}-\ref{a42}), we see that the scalar function
\begin{eqnarray}\label{a15}
&\Omega^0=\mu^0+\mathbf{v}^{0}_{s}\cdot \mathbf{v}_{n},\nonumber\\
 &\Omega^a=\mu^a+\mathbf{v}^{a}_{s}\cdot \mathbf{v}_{n}.
\end{eqnarray}
From Eq.(\ref{a151-2}), we can see the superfluid velocity of U(1) part  $\mathbf{v}_{s}^{0}$ is still irrotational, i.e., $\nabla \times \mathbf{v}_{s}^{0}=0$.


If the energy density $\epsilon$ or free energy $f$ density is invariant under a general U(N) (global) rotation, e.g., $\delta n^a=f^{a}_{\ bc}n^b\delta\theta^c$, $\delta \mu^a=f^{a}_{\ bc}\mu^b\delta\theta^c$, $\delta h^a=f^{a}_{\ bc}h^b\delta\theta^c$ and $\delta \mathbf{v}^{a}_{s}=f^{a}_{\ bc}\mathbf{v}^{b}_{s}\delta\theta^c$, the last term for $n^a$  in Eq.(\ref{a151-1}) is zero, i.e.,
\begin{eqnarray}\label{invariance}%
f^{a}_{\ bc}[\mu^b n^c+h^{b}_{i}\textrm{v}^{c}_{si}]\equiv0.
\end{eqnarray}%
It means that $n^a$ satisfies a continuity equation and generalized spin (magnetization)  $n^a$ is conserved in a U(N) invariant superfluid. Eqs.(\ref{a151}-\ref{a151-5}) can be applied to a system with arbitrary symmetry group. We see that there are many superfluid velocities $\mathbf{v}_{s}^a$ in  general.

In the following, we will see that for a U(N) invariant superfluid, we can use a single superfluid velocity $\mathbf{v}_s$ in the hydrodynamic description. In this case, the hydrodynamic equation is simplified greatly.
Now for a U(N) invariant superfluid, we take a specific form for free energy density as
\begin{eqnarray}\label{a17}
f &=f_{0}(T,n,n^a)+f_{\rm Landau}+f_{\rm kinetic},\nonumber\\
f_{\rm Landau} &=\frac{\alpha (\nabla n^a)^2}{2},\nonumber\\
f_{\rm kinetic} &=\frac{mn\mathbf{v}^{2}_n}{2}+\frac{mn_s(\tilde{\mathbf{v}}^{0}_{s}-\mathbf{v}_n)^2}{2},
\end{eqnarray}%
where $\tilde{\mathbf{v}}^{0}_{s}=\mathbf{v}^{0}_{s}+({n^a}/{n})\mathbf{v}^{a}_{s}$ is an effective superfluid velocities.  $f_{0}(T,n,n^a)$ is the free energy density when superfluid is in equilibrium, i.e., $\mathbf{v}_n=\mathbf{\tilde{v}}_s=\nabla n^a=0$.  $f_{\rm Landau}$ is the energy of magnetization distortion (gradient energy) in ferromagnetic system. The remaining part $f_{\rm kinetic}$ is kinetic energy when $\mathbf{v}_n\neq0$ or $\mathbf{\tilde{v}}_s\neq0$.
Here we should identify $\tilde{\mathbf{v}}^{0}_{s}$ with $\mathbf{v}_s$ in the main text. We see that the superfluid velocity $\tilde{\mathbf{v}}^{0}_{s}$ is mixture of superflow motion and spin rotation (magnetization distortion). Using Eq.(\ref{a2-1}) and Eq.(\ref{a91}), $f_{\rm Landau}$ can be rewritten as
\begin{eqnarray}\label{a192}
\nabla n^a &=f^{a}_{\ bc} n^{b} \mathbf{v}^{c}_{s},\nonumber\\
f_{\rm Landau} &=\frac{\alpha (\nabla n^a)^2}{2}=\frac{\alpha}{2}f^{a}_{\ bc}f^{a}_{\ ef}n^bn^e\textrm{v}^{c}_{si}\textrm{v}^{f}_{si}.
\end{eqnarray}

In the following, we will label all the physical quantities with tilde, which would be identified with the corresponding ones in the main text.
By Eq.(\ref{a17}) and Eq.(\ref{a192}), and comparing two thermodynamic relations Eqs.(\ref{566-1}-\ref{566-5}), Eqs.(\ref{a101-1}-\ref{a101-5}),  the relations between old and new ones are :
\begin{eqnarray}
&h^0_{i}=\tilde{h}^0_{i}=mn_s(\tilde{\textrm{v}}^{0}_{si}-\textrm{v}_{ni}),\label{a193-1}\\
&h^a_{i}=\alpha f^{d}_{\ ba}f^{d}_{\ ef}n^bn^e \textrm{v}^{f}_{si}+\frac{n^a}{n}h^{0}_{i}\equiv\hat{h}^a_{i}+\frac{n^a}{n}h^{0}_{i},\\
&\mu^0=\tilde{\mu}^{0}-\frac{n^a\textrm{v}^{a}_{sj}}{n^2}h^{0}_{j},\\
&\mu^a=\alpha f^{d}_{\ ac}f^{d}_{\ ef}n^e\textrm{v}^{c}_{sj} \textrm{v}^{f}_{sj}+\frac{\partial f_0}{\partial n^a}+\frac{\textrm{v}^{a}_{sj}}{n}h^{0}_{j}
\equiv\hat{\mu}^a+\frac{\textrm{v}^{a}_{sj}}{n}h^{0}_{j},\\
&\tilde{\mu}^a=-\alpha\nabla_j [\nabla_j n^a]+\frac{\partial f_0}{\partial n^a}=-\alpha\nabla_j [f^{a}_{\ bc} n^{b} \textrm{v}^{c}_{sj}]+\frac{\partial f_0}{\partial n^a}.\label{a193-5}
\end{eqnarray}%
To show that our assumption of the effective superfluid velocity $\tilde{\mathbf{v}}^{0}_{s}=\mathbf{v}^{0}_{s}+({n^a}/{n})\mathbf{v}^{a}_{s}$ satisfies the Eq.(\ref{continuity11}) that was derive in the main text, we calculate the equation of motion of $\tilde{\mathbf{v}}^{0}_{s}$ and show that they are consistent. From Eq.(\ref{a151}-\ref{a151-3}),  we get
\begin{eqnarray}\label{a201}
\frac{\partial \tilde{\textrm{v}}^{0}_{si}}{\partial t}&=\frac{\partial \textrm{v}^{0}_{si}}{\partial t}+\frac{n^a}{n}\frac{\partial \textrm{v}^{a}_{si}}{\partial t}+\textrm{v}^{a}_{si}\partial_t(\frac{n^a}{n})\nonumber\\
&=-\nabla_i[\mu^0+\textbf{v}_n\cdot \tilde{\textbf{v}}^{0}_{s}]-\frac{n^a \nabla_i\mu^a}{n}+\textbf{v}^{a}_{s}\cdot \textbf{v}_n \nabla_i(n^a/n)\nonumber\\
&~~~~+\textrm{v}^{a}_{si}\partial_t (n^a/n)+\frac{n^a}{n}f^{a}_{\ bc}[\mu^b+\textbf{v}^{b}_{s}\cdot \textbf{v}_n]v^{c}_{si}\nonumber\\
&=-\nabla_i[\mu^0+\textbf{v}_n\cdot \tilde{\textbf{v}}^{0}_{s}]-\frac{n^a \nabla_i\mu^a}{n}\nonumber\\
&~~~~+\textrm{v}^{a}_{sj}\textrm{v}_{nj}\nabla_i(n^a/n)-\textrm{v}^{a}_{si}\textrm{v}_{nj}\nabla_j(n^a/n)\nonumber\\
&~~~~+\frac{-n\textrm{v}^{a}_{si}\nabla_jh^{a}_{j}+n^a\textrm{v}^{a}_{si}\nabla_jh^{0}_{j}}{n^2}+\frac{n^a}{n}f^{a}_{\ bc}[\mu^b+\textbf{v}^{b}_{s}\cdot \textbf{v}_n]\textrm{v}^{c}_{si}\nonumber\\
&=-\nabla_i[\mu^0+\textbf{v}_n\cdot \tilde{\textbf{v}}^{0}_{s}]-\frac{n^a \nabla_i\mu^a}{n}\nonumber\\
&~~~~+\textrm{v}_{nj}[\nabla_i(n^a\textrm{v}^{a}_{sj}/n)-\nabla_{j}(n^a\textrm{v}^{a}_{si}/n)]\nonumber\\
&~~~~+\frac{-n\textrm{v}^{a}_{si}\nabla_jh^{a}_{j}+n^a\textrm{v}^{a}_{si}\nabla_jh^{0}_{j}}{n^2}+\frac{h^{a}_{j}}{n}[\nabla_i\textrm{v}^a_{sj}-\nabla_j\textrm{v}^{a}_{si}] \ \ \  \nonumber\\
&=-\nabla_i[\mu^0+\textbf{v}_n\cdot \tilde{\textbf{v}}^{0}_{s}]-\frac{n^a \nabla_i\mu^a}{n}
+\textrm{v}_{nj}[\nabla_i\tilde{\textrm{v}}^{0}_{sj}-\nabla_{j}\tilde{\textrm{v}}^{0}_{si}]\nonumber\\
&~~~~+\frac{-n\textrm{v}^{a}_{si}\nabla_jh^{a}_{j}+n^a\textrm{v}^{a}_{si}\nabla_jh^{0}_{j}}{n^2}+\frac{h^{a}_{j}}{n}[\nabla_i\textrm{v}^a_{sj}-\nabla_j\textrm{v}^{a}_{si}]\ \ \  \nonumber\\
&=-\nabla_i[\tilde{\mu}^0+\textbf{v}_n\cdot \tilde{\textbf{v}}^{0}_{s}]-\frac{n^a \nabla_i\tilde{\mu}^a}{n}-[(\nabla\times \tilde{\textbf{v}}^{0}_{s})\times \textbf{v}_n]_i\nonumber\\
&~~~~-\frac{1}{n}[(\nabla\times \tilde{\textbf{v}}^{0}_{s})\times \textbf{h}^0]_i\nonumber\\
&~~~~-\frac{\textrm{v}^{a}_{si}\nabla_j \hat{h}^{a}_{j}}{n}+\frac{\hat{h}^{a}_{j}}{n}[\nabla_i\textrm{v}^a_{sj}-\nabla_j\textrm{v}^{a}_{si}]-\frac{n^a\nabla_i[\hat{\mu}^a-\tilde{\mu}^a]}{n}.
\end{eqnarray}%
In the above derivation, we have used the zero-curvature condition Eq.(\ref{zerocurvature}),  the U(N)  invariance of the energy density, Eq.(\ref{invariance}), Eqs.(\ref{a193-1}-\ref{a193-5}) and the fact that $\nabla \times \tilde{\mathbf{v}}^{0}_{s}=\nabla\times (n^a\mathbf{v}^{a}_{s}/n)$ (due to $\nabla \times \mathbf{v}_{s}^{0}=0$).

Similarly, the equation for $g_i$ is
\begin{eqnarray}\label{a211}
&\frac{\partial g_{i}}{\partial t}=-\nabla_j\pi_{ij}=-\nabla_j[\tilde{p}\delta_{ij}+g_{j}\textrm{v}_{ni}+\tilde{\textrm{v}}^{0}_{sj}h^{0}_{i}]\nonumber\\
&~~~~~~~~~~-\textrm{v}^{a}_{si}\nabla_j\hat{h}^{a}_{j}+\hat{h}^{a}_{j}[\nabla_i\textrm{v}^a_{sj}-\nabla_j\textrm{v}^{a}_{si}]-n^a\nabla_i[\hat{\mu}^a-\tilde{\mu}^a]
\end{eqnarray}%

If the last line in Eq.(\ref{a201}) or Eq.(\ref{a211}) is zero, both equations would reduce to the equations of $\textbf{v}_s$ and $g_i$ in the main text [see Eqs.(\ref{continuity})-Eq.(\ref{continuity11})]. The last line is proportional to
\begin{eqnarray}\label{a21}
&-\textrm{v}^{a}_{si}\nabla_j\hat{h}^{a}_{j}+\hat{h}^{a}_{j}[\nabla_i\textrm{v}^a_{sj}-\nabla_j\textrm{v}^{a}_{si}]-n^a\nabla_i[\hat{\mu}^a-\tilde{\mu}^a]\nonumber\\
=&-\alpha f^{d}_{\ ba}f^{d}_{\ ef}n^bn^e(\nabla_j \textrm{v}^{f}_{sj})\textrm{v}^{a}_{si} -\alpha f^{a}_{\ bc}f^{b}_{\ ef}n^an^e (\nabla_j\textrm{v}^{c}_{sj})\textrm{v}^{f}_{si}\ \ \ \ \ \nonumber\\
=&-\alpha f^{d}_{\ ba}f^{d}_{\ ef}n^bn^e(\nabla_j \textrm{v}^{f}_{sj})\textrm{v}^{a}_{si}-\alpha f^{d}_{\ ae}f^{d}_{\ bf}n^bn^e (\nabla_j\textrm{v}^{f}_{sj})\textrm{v}^{a}_{si}-\alpha f^{d}_{\ eb}f^{d}_{\ af}n^bn^e (\nabla_j\textrm{v}^{f}_{sj})\textrm{v}^{a}_{si}\nonumber\\
=&-\alpha[f^{d}_{\ ba}f^{d}_{\ ef}+f^{d}_{\ ae}f^{d}_{\ bf}+f^{d}_{\ eb}f^{d}_{\ af}]n^bn^e (\nabla_j\textrm{v}^{f}_{sj})\textrm{v}^{a}_{si}\nonumber\\
=&0.
\end{eqnarray}%
In the above calculation, the zero-curvature condition Eq.(\ref{zerocurvature}) and Jacobi identity Eq.(\ref{Jacobi}) in the main text are used. In addition, using Jacobi identity Eq.(\ref{Jacobi}), we can prove that
\begin{eqnarray}\label{a231}
&[f^{d}_{ba}f^{d}_{ef}f^{a}_{kl}+f^{d}_{bk}f^{d}_{af}f^{a}_{el}]n^en^b\textrm{v}^f\textrm{v}^l\textrm{v}^{k}_{i}\nonumber\\
=&[f^{a}_{dl}f^{a}_{bk}+f^{a}_{db}f^{a}_{kl}]f^{d}_{ef}n^en^b\textrm{v}^f\textrm{v}^l\textrm{v}^{k}_{i} \nonumber\\
=&-[f^{a}_{dk}f^{a}_{lb}]f^{d}_{ef}n^en^b\textrm{v}^f\textrm{v}^l\textrm{v}^{k}_{i}
\nonumber\\
=&[f^{a}_{dk}f^{a}_{lb}]f^{d}_{fe}n^en^b\textrm{v}^f\textrm{v}^l\textrm{v}^{k}_{i}\nonumber\\
=&\frac{1}{2}[f^{k}_{ad}f^{a}_{lb}f^{d}_{fe}+f^{k}_{da}f^{a}_{fe}f^{d}_{lb}]n^en^b\textrm{v}^f\textrm{v}^l\textrm{v}^{k}_{i}\nonumber\\
=&\frac{1}{4}[f^{k}_{ad}f^{a}_{lb}f^{d}_{fe}+f^{k}_{da}f^{a}_{fe}f^{d}_{lb}+f^{k}_{da}f^{d}_{lb}f^{a}_{fe}+f^{k}_{da}f^{d}_{fe}f^{a}_{lb}]n^en^b\textrm{v}^f\textrm{v}^l\textrm{v}^{k}_{i}\nonumber\\
=&0,
\end{eqnarray}%
which is also used in Eq.(\ref{a21}).

Finally, one can also show that the other remaining equations for $n$, $n^a$ and $s$ in Eqs.(\ref{a151}-\ref{a151-5}) are the same as we have derived in the main text. Therefore, we conclude that under an assumption of free energy Eq.(\ref{a17}), the hydrodynamic equation of U(N) invariant superfluid Eqs.(\ref{continuity})-(\ref{continuity11}) can be obtained from the general hydrodynamic equation Eqs.(\ref{a151}-\ref{a151-5}) with arbitrary internal symmetry group.

\subsection{The origin of breaking of irrotationality  of superfluid velocity $\mathbf{v}_s$}
One may wonder why the superfluid velocity $\mathbf{v}_s$  does not satisfy the irrotaional condition in U(N) invariant superfluid. This is because of the coupling between the superfluid motions of U(1) part and SU(N) part. The unique superfluid velocity $\mathbf{v}_s$ is identified as
\begin{eqnarray}\label{a242}
\mathbf{v}_s\equiv\tilde{\mathbf{v}}^{0}_s\equiv \mathbf{v}^{0}_s+\frac{n^a\mathbf{v}^a}{n},
\end{eqnarray}
which is  a mixture of several superfluid velocities.
 Now it is assumed that there is a variation of order parameter $\psi$ (or field operator) under a rotation
($\psi\rightarrow e^{i\delta \theta^0 T^0+i\delta \theta^b T^b}\psi$, $\psi^\dag\rightarrow \psi^\dag e^{-i\delta \theta^0 T^0-i\delta \theta^b T^b}$), i.e.,
\begin{eqnarray}
  \delta \psi=e^{i\delta \theta^0 T^0+i\delta \theta^b T^b}\psi-\psi=(iT^0\delta\theta^0+iT^a\delta\theta^a)\psi.
\end{eqnarray}
 By Eq.(\ref{a9}) and Eq.(\ref{a2-1}), such a rotation would induce a momentum density $\mathbf{g}$
 \begin{eqnarray}
  \mathbf{g}=n\mathbf{v}^{0}_s+n^a\mathbf{v}^{a}_s=n(\mathbf{v}^{0}_s+\frac{n^a\mathbf{v}^{a}_s}{n})\equiv n\mathbf{\tilde{v}}^{0}_s.
\end{eqnarray}
Then we find that the effective superfluid velocity
 \begin{eqnarray}
\mathbf{v}_s\equiv\tilde{\mathbf{v}}^{0}_s\equiv \mathbf{v}^{0}_s+\frac{n^a\mathbf{v}^{a}_{s}}{n},
\end{eqnarray}
is  a mixture of  the velocities of the U(1) part and SU(N) part.
Because the  superfluid velocity $\mathbf{v}_s$ can not be written as a derivative of a scalar function, then the irrotational condition ($\nabla \times \mathbf{v}_s=0$) can not be satisfied in general.

At last, we notice  that the correct commutation relations [Eqs.(\ref{38})-(\ref{38-5})] between $\mathbf{v}_s$ and $n$, $n^a$, $s$, $\textbf{g}$ in the main text can be also obtained by using the fundamental commutation relation Eq.(\ref{a123})-(\ref{a123-5}) and the definition $\mathbf{v}_s\equiv\mathbf{v}^{0}_s+\frac{n^a\mathbf{v}^{a}_{s}}{n}$. Such a method provides another derivation of the hydrodynamic Eqs.(\ref{continuity}-\ref{continuity11}) in the main text.

\section{References}

\end{document}